%% file: main.tex
\documentclass[sigconf]{acmart}

\AtBeginDocument{%
  \providecommand\BibTeX{{%
    \normalfont B\kern-0.5em{\scshape i\kern-0.25em b}\kern-0.8em\TeX}}}

\setcopyright{acmcopyright}
\copyrightyear{2024}
\acmYear{2024}
\acmDOI{XXXXXXX.XXXXXXX}

\acmConference[DIS '24]{Make sure to enter the correct
  conference title from your rights confirmation email}{July 1--5,
  2024}{Coopenhagen, DK}
\acmBooktitle{ACM Designing Interactive Systems (DIS
’24),
 July 1--5, 2024, Coopenhagen, DK} 
\acmPrice{15.00}
\acmISBN{978-1-4503-XXXX-X/18/06}

\input{utils.tex}
\input{aliases.tex}

\begin{document}

\title{``For Us By Us'': Intentionally Designing Technology for Lived Black Experiences}

\author{Lisa Egede}
\affiliation{%
 \institution{Carnegie Mellon University}
 \city{Pittsburgh}
 \state{PA}
 \country{USA}}
 \email{legede@andrew.cmu.edu}

\author{Leslie Coney}
\affiliation{%
 \institution{University of Washington}
 \city{Seattle}
 \state{WA}
 \country{USA}}
\email{lesconey@uw.edu}

\author{Brittany Johnson}
\affiliation{%
  \institution{George Mason}
  \city{Fairfax}
  \state{VA}
  \country{USA}}
\email{johnsonb@gmu.edu}

\author{Christina N. Harrington}
\affiliation{%
  \institution{Carnegie Mellon University}
  \city{Pittsburgh}
  \state{PA}
  \country{USA}}
\email{charring@andrew.cmu.edu}

\author{Denae Ford}
\affiliation{%
  \institution{Microsoft Research}
  \city{Redmond}
  \state{WA}
  \country{USA}}
\email{denae@microsoft.com}

\renewcommand{\shortauthors}{Egede, L.  Coney, L., Johnson, B., Harrington, C., Ford, D.}

\begin{abstract}
HCI research to date has only scratched the surface of the unique approaches racially minoritized communities take to building, designing, and using technology systems. While there has been an increase in understanding how people across racial groups create community across different platforms, there is still a lack of studies that explicitly center on how Black technologists design with and for their own communities. In this paper, we present findings from a series of semi-structured interviews with Black technologists who have used, created, or curated resources to support lived Black experiences. From their experiences, we find a multifaceted approach to design as a means of survival, to stay connected, for cultural significance, and to bask in celebratory joy. Further, we provide considerations that emphasize the need for centering lived Black experiences in design and share approaches that can empower the broader research community to conduct further inquiries into design focused on those in the margins. 

\end{abstract}



\maketitle

\input{Sections/1_Introduction}

\input{Sections/2_RelatedWork}

\input{Sections/3_Method}

\input{Sections/4_Findings}
\input{Sections/5_Discussion}

\input{Sections/6_ConclusionandFutureWork}

\begin{acks}
We would like to thank the tech entrepreneurs and founders for participating in our study and partnering with us on this research project. We would also like to thank Geoff Kaufman for his valuable feedback on this work. 
\end{acks}

\bibliographystyle{ACM-Reference-Format}
\bibliography{references}

\appendix
\newpage

\end{document}

%% file: utils.tex
\usepackage{hyperref}

\usepackage{verbatim}

\usepackage{xspace}
\xspaceaddexceptions{]\}}

\usepackage{xcolor}
\usepackage{lscape}
\usepackage{colortbl}

\usepackage{array}
\usepackage{wrapfig}
\usepackage{multirow}
\usepackage{multicol}
\usepackage{tabularx}

\usepackage{lookup}
\input{lookup_table}

\usepackage{fontawesome5}

\newcommand{\quo}[2]{ 
	\vspace{-0.15cm}
	\def\FrameCommand{%
		\hspace{0pt}%
		{\color{PAblue}\vrule width 2pt}%
		{\color{white}\vrule width 2pt}%
		\colorbox{white}
	}%
	\MakeFramed{\advance\hsize-\width\FrameRestore}%
	\noindent\hspace{-4.55pt}%
	\begin{adjustwidth}{}{0pt}
		\vspace{-2pt}%
		``\emph{#1}'' (\lookupGet{#2})
		\vspace{-3pt}
	\end{adjustwidth}\endMakeFramed%
	\vspace{-0.15cm}
}

\newcommand{\inlinequote}[1]{``\emph{#1}''}

\newcommand{\blockquo}[2]{
\begin{quote}
``{#1}'' -\participant{#2}
\end{quote} }

%% file: lookup_table.tex
\lookupPut{P1}{1}
\lookupPut{P2}{2}
\lookupPut{P3}{NB1} %
\lookupPut{P4}{3} 
\lookupPut{P5}{4}
\lookupPut{P6}{5}
\lookupPut{P7}{NB2} %
\lookupPut{P8}{6}
\lookupPut{P9}{7}
\lookupPut{P10}{8}
\lookupPut{P11}{9}
\lookupPut{P12}{10}
\lookupPut{P13}{11}
\lookupPut{P14}{NB3} %
\lookupPut{P15}{12}
\lookupPut{P16}{13}
\lookupPut{P17}{14}

%% file: aliases.tex
\usepackage{pgfplots}
\pgfplotsset{width=10cm,compat=1.9}

\usepackage{stackengine}

\usepackage{xcolor}
\definecolor{darkgreen}{rgb}{0.05,0.5,0.05}

\newcommand{\RQA}{ What is technology’s role in supporting lived Black experiences?\xspace}
\newcommand{\RQB}{How do technologists design for lived Black experiences?\xspace}
\newcommand{\RQC}{What are the experiences of the technologists  building these projects?\xspace}

\newcommand{\LBE}{lived Black experience\xspace}

%% file: Sections/1_Introduction.tex
\section{Introduction}
HCI has begun to engage with ontologies of culture and identity such as race, nationality, or ethnicity as actual constructs that are both embedded into and influencing the way many experience technology~\cite{cunningham2022grounds, himmelsbach2019we}. From the way we think about methods and who is included in the design of things, to the ways that interactive devices impact certain communities, considerations of cultural experiences in design have become a critical aspect of the discourse around transformative and inclusive approaches to design. Many of the conversations considering technology experiences among racially minoritized groups focus on equitable and culturally tailored interactions. Recent work has even explored how we might engage racial identity as a ``cultural ethos" due to the unique ways that certain groups perceive and experience technological systems~\cite{winchester2020black, bray2021speculative, winchester2019engaging}. 
We have seen changes in design practices as HCI has begun to look to cultural identity and the distinctiveness of cultural likeness as ways to create more inclusive products and online communities. Engaging with both identity and likeness as the interactions individuals have with others intends to create products that are both relatable and reflective of users' basic psychological needs, and speak to their desire for community and sense of belonging~\cite{klassen2021more, musgrave2022experiences}. Researchers and practitioners alike are articulating \textit{why} racially minoritized groups like Black Americans should not only be considered in how things are designed but should be the architects of these design decisions~\cite{rankin2020seat, rankin2019exploring, harrington2021eliciting}. This is particularly fitting as a response to the Black art renaissance moment that has emerged in the last few years following the COVID-19 pandemic, civil rights protests, and other social movements~\cite{howell_mills_2020, ray2017evolution, rolling2023resilience}. Influenced by this, marginalized groups are finding ways to express themselves through media, art, literature, design, and the technology they use~\cite{rolling2023resilience}. 

Communities of HCI research and practice have long honored the value of culturally tailoring technology experiences to make them personally relevant and increase positive attitudes towards said interventions~\cite{oleary2020community, noar2007does}. Within the realm of culturally tailored tech, there is a growing call for such tailoring to be led by those who are directly impacted by technology. For example, the phrase \textit{Nothing About Us Without Us} has its earliest association as a slogan that advocates for disabled persons to be included as active participants in issues involving disability access~\cite{spiel2020nothing}. \textit{`Nothing About Us Without Us'}~\cite{spiel2020nothing}, similar to the popular \textit{`For Us By Us'}\footnote[1]{For Us By Us or F.U.B.U. is a popular clothing brand founded in 1992 https://www.theroot.com/for-us-by-us-by-any-means-necessary-daymond-john-tell-1845810402.} campaign, suggests that systems that are intended for a group should be developed by-- or at the very least include- that group in the process. Its sentiment has since expanded across HCI to ameliorate issues of other marginalized groups such as children, older adults, and racially minoritized persons~\cite{harbord2021indigenous, kane2014collab, costanza2020design, spiel2020nothing}. Design approaches such as co-design, participatory design, or even community-based participatory methods have been popularly associated with such cultural tailoring and inclusion, yet an important consideration of design that advocates for the inclusion of any group is the many ways that this group contextualizes their experience. Although no user group can be considered a monolith, we posit that there is value in the exploration of identity-specific aspects of approaching design projects and technologies that are meant to be inclusive, safe, and culturally aware.  

Contextualizing designing for lived Black experiences has been touched upon in prior literature both in and out of HCI and design~\cite{harrington2021designing, dancy2021ai, cunningham2022grounds, bray2021speculative}. Dancy positions that for us to understand the intersection of AI and Blackness, we must understand Blackness beyond identity and category and instead understand it as \textit{``a sign, symbol, or metaphor that represents something''}~\cite{dancy2021ai, fanon2008black}. Tying into this idea Cunningham and colleagues position that design must do more than associate Blackness with problem-solution thinking and ultimately solutionism~\cite{cunningham2022grounds}. Winchester argues that Blackness and by extension `Black-centered' design may be an actual ethos and framework that pushes inclusion by considering the needs of one of the most marginalized groups in the U.S.~\cite{winchester2020black}. Black digital studies researcher André Brock's work exploring Blackness in internet culture suggests that \textit{``Blackness has expertly utilized the internetwork’s capacity for discourse to build out a social, cultural, [and] racial identity"} and is something merely performed when mediated by technology~\cite{brock2020distributed}. What then does it mean to design for the '\LBE' in a way that doesn't replicate systemic oppression, comparisons to whiteness, or ontological otherness? We argue that designing for their experiences also means understanding potential facets of desired interactions with and perspectives of technology's role in Black lives. In defining what is meant by the ``\LBE'', we build off of the foundation other HCI scholars have established by looking at Blackness as more than mere race but as a social construct that has othered cultural groups against a standard norm or oppressive structures.

In this paper, we present a qualitative study as a part of a larger research agenda to operationalize what is meant by ``the Black experience'' in the design of products and online spaces. Our intent is to identify projects that fulfill this through the inclusion of Black communities, and ultimately make this work more visible. We build on prior literature that has introduced ontologies of Blackness as more than a problem to be solved or a comparison group against a standard of whiteness~\cite{cunningham2022grounds}, but as a cultural ethos that has unique orientations that may inform design~\cite{winchester2019engaging, bray2021speculative}. As a part of our qualitative research study, we interviewed 17 creators, technologists, and founders from across the United States and Canada to highlight technology projects, products, and apps intended to amplify lived Black experiences. Specifically, we wanted to move from speculation to understanding how technologists who have built these systems position the work they've done in this space. We first contextualize lived Black experiences from a sample of Black and non-Black technologists, creators, and curators who have identified a need for a product or service and brought it to life. Specifically, we sought to address the following research questions: 

\begin{description}
\setlength{\itemsep}{0pt}
\setlength{\parskip}{0pt}
    \item[RQ1] \RQA
    \item[RQ2] \RQB
    \item[RQ3] \RQC
\end{description}

The research questions posed tie into themes around understanding support needs, design goals, and technologists' perspectives with the overall goal being to develop a larger picture of what it means to design for lived Black experiences. We present an analysis of 17 interviews spanning 17 products and online communities. We position our work among literature that contextualizes culture and identity against the backdrop of technology experiences and examines visibility and inclusion among tech creators. Our work presents unique opportunities for the DIS community to consider when designing for racially minoritized groups. Specifically, findings from this study contribute a taxonomy outlining how people have designed tech for lived Black experiences, which will then inform how designers engage with these communities. To promote equitable and long-term engagement with Black technologists, we provide recommendations for the broader DIS community to consider when it comes to how they amplify the leaders of projects supporting historically marginalized groups.

%% file: Sections/2_RelatedWork.tex
\section{Related Work}

 Contextualizing culturally aware technology can be thought of as a key component in inclusive design.  While prior research has started to situate ``Blackness'' as a design ethos~\cite{bray2022radical, winchester2018afrofuturism, bray2021speculative,harrington2021designing, cunningham2022cost}, there is a need for more work that centers the design needs of Black individuals or ethnographic work that understands how these groups are designing for their communities. We situate our research by providing an overview of current work which considers the lack of diversity in tech spaces, the need to incorporate Black experiences into technological systems, and the design of products and online communities that are designed for Black communities by Black technologists. In doing so we will present works related to the design of products and online communities ``For Us By Us'' .

\subsection{Lack of Diversity in Tech Spaces and its Impacts}

Disparities in the design space are prevalent both in who is responsible for the design of technological systems and who is the focus of these designs~\cite{linxen2021weird}. This historical under-representation has resulted in racism being embedded in everyday systems thus rendering racial disparities in design as ``the norm," a topic that is touched on in Ogbonnaya et al's Critical Race Theory in HCI~\cite{ogbonnaya2020critical}. Ogbonnaya and colleagues push for the HCI community to acknowledge the role race plays in HCI and to center the experiences of historically excluded groups in design more broadly~\cite{ogbonnaya2020critical, himmelsbach2019we}.  

As it stands, the lack of tech-centered design that considers Black users can be attributed to the low number of Black people in tech as shown by the 2014 U.S. Equal Employment Opportunity Commission report~\cite{equale}. According to this report, 7.4\% of people employed in the tech sector are Black, and only 2\% made up employment positions on an executive level~\cite{equale}. The lower percentage of tech and design employment among Black Americans is both influenced by and impacts feelings of belonging,~\cite{farnsworth2016diversity,  alexander2016african, kang_frankel_2015, alfrey2022diversity}, and skews the focus of design to prioritize the experience of some groups over others. 

When looking at whose experiences are centered in the creation of technological systems, it is important to consider the intentional inclusion of diverse perspectives, as touched on in Linxen et al's work discussing the overrepresentation of WEIRD (Western, Educated, Industrialized, Rich, and Democratic)~\cite{linxen2021weird} participants in CHI paper samples. Similar to prior work around the negative impacts of sample bias in HCI research~\cite{dell2012yours, offenwanger2021diagnosing}, Linxen's work highlights the impact that including both WEIRD and non-WEIRD perspectives can have on the design of tech systems and the communities are more susceptible to its negative impacts~\cite{linxen2021weird}. Additionally, there is a need for diverse representation amongst tech creators and designers themselves. Understanding what it means to center underrepresented individuals in the design process from the perspective of Black creators and technologists may work toward a more inclusive design process. 

\subsection{Incorporating the `Black Experience' into the Design of Technological Systems}

``Blackness'' as a concept is most commonly explored in African American studies, sociology, and racial studies ~\cite{maylor2014meaning, sudbury2001re, brunsma2002does}. Equity and social justice education professor Uvanney Maylor's work draws on the existing contention around the meaning of Blackness across various contexts and found that it was insufficient to capture the experiences of all Black people under a single terminology~\cite{maylor2014meaning}, emphasizing this idea that \textit{‘Black’ identities are ‘constantly [being] redefined in the light of shifting public discourse and political necessities’~\cite{sudbury2001re}}. The unclear definition around the true meaning of ``Blackness'' can be attributed to the unique experiences of the Black communities across different contexts, including geographically, socially, and financially~\cite{drake1980anthropology, maylor2014meaning, sudbury2001re, hall1993black, brunsma2002does}. Given these complexities, attempts to reference a single Black experience when creating technology or tools targeting Black communities often fall short~\cite{moten2008case, hrabovsky2013concept}, which can explain why emerging HCI research has sought to better engage with identity-specific problem cases through design.

Prior work in HCI has drawn on the importance of engaging with culture and identity to inform the design of technology, with newer literature directly engaging constructs of racial identity and more explicitly ``the Black experience'' in design~\cite{harrington2021designing, ogbonnaya2020critical}. Across the literature in HCI and design, research scholars such as Rankin, Cunningham, Bray, Klassen, Bosely, and Harrington discuss concepts such as ontological or speculative ``Blackness'' or the ``Black experience'' as a way to expand both the impact of and access to design~\cite{rankin2020intersectionalexperiences, cunningham2022cost, cunningham2022grounds, bray2021speculative, klassen2021more, harrington2021designing}. In the \textit{Interactions} perspective paper, Harrington et al discuss the value of academic research and design practices, calling for the broader HCI community to engage with the different dimensions that embody the Black experience~\cite{harrington2021designing}. In particular, there is a need to expand the consideration of this experience to include the ``joy-focused'' parts of Black culture~\cite{harrington2021designing, harrington2021eliciting, bray2022radical}. Prior work points to how an inaccurate perception of community tropes or stereotypes may unintentionally perpetuate existing harms~\cite{tran2019gets, cunningham2022grounds}.  
O'Leary et al's work highlights this orientation in the 2019 study on the community group \textit{``Africatown''} and their efforts to combat displacement in a historically Black neighborhood as a result of harmful design practices by designers and developers~\cite{tran2019gets}. By drawing on the real-world impacts that conventional design practices can have, this work supports the claim that such ``solution-based'' approaches with minimal focus on the community's actual lived experiences can have harmful long-term impacts on communities~\cite{tran2019gets, cunningham2022grounds}. 

In attempts to avoid these harms, several HCI researchers have shown that amplifying the cultural heritage of Black and other racially-minoritized communities may shed light on their experiences in a way that can inform design~\cite{tran2019gets, grimes2008celebratory, informing2018, martinhammond2022bridging, kim2022designing}. In ``Celebratory Technology"~\cite{grimes2008celebratory}, Grimes highlights how looking at the communal practices of eating and cooking may support designing nutrition solutions among lower-income Black Americans. When looking at prior work that highlights the value of designing for the everyday experiences of racially minoritized communities, To et al emphasize the importance of centering joy, pleasure, and cultural heritage when designing for BIPOC communities, stating that there is no single framework for designing for these communities, as the process is ``constantly emerging'' ~\cite{to2023flourishing}. Similarly, Harrington discusses how engaging Black older adults in community-based design can push design to include more holistic accounts of Black communities that speak to strengths instead of deficits only~\cite{forgottenmargins2020}. Rankin and Irish discuss how centering Black woman designers in game design and embracing their knowledge and lived experiences can be transformational~\cite{rankin2020seat}. Findings from their study emphasized that there is a need to re-frame Black designers as ``agents of knowledge'', rather than ``objects of knowledge'' for academic inquiry; a reframing that emphasizes valuing the lived experiences of Black women and the impact their viewpoints hold in the design process~\cite{rankin2020seat}. Similarly, Erete et al discuss how ``counter-narratives'' may be a way to create more inclusive and representative spaces in computing for Black and Hispanic women~\cite{erete2021applying}. 

As this research area grows, we find that there is a need to operationalize how designers conceptualize ontological Blackness~\cite{cunningham2022grounds} or the lived experience of Black individuals if for no other reason than to mitigate emerging bias and harms caused by color-blind and negligent systems~\cite{winchester2020black}. Thus far, HCI researchers have laid a foundation for the integration of ``Blackness" in design as an ethos for healing~\cite{bosley2022healing}, inclusive technological innovation~\cite{winchester2018afrofuturism, winchester2020black}, and world building~\cite{bray2022radical}. Scholars such as Woodrow Winchester argue that: \textit{``Black-Centered Design''} approaches offer a framework by which the nuanced complexities of the Black identity can act as an ethos for creating more equitable and just emerging technological solutions''~\cite{winchester2019engaging}. In considering how design may practically apply dimensions of culture and heritage when designing for Black communities, works by both Winchester and Bray push for designers to engage with frameworks such as Afrofuturism as an attempt to center Black experiences by connecting both the past, present, and future of Black communities~\cite{winchester2018afrofuturism, bray2021speculative}. Despite the existing body of work that considers the value of bridging culture, heritage, and design or more specifically incorporating Black experiences in design, few papers have empirically explored and explicitly contextualized what this means. 

\subsection{Designing Products and Online Communities \textit{``For Us By Us''}}
Recently, we have begun to see racially minoritized communities re-center their experiences through design as a means to combat the misdirection of design efforts by creating products and online communities that capture a more holistic narrative~\cite{harbord2021indigenous}. Hemayssi et al's work on designing user interfaces to center Arabic culture found that having experts whose lived experiences are situated in the design context can have meaningful impacts on the overall usefulness of the interface~\cite{hemayssi2005designing}. Specifically, language and culture play a major role in how users contextualize their experiences~\cite{hemayssi2005designing}. This work further emphasizes the importance of including stakeholders whose identities are situated in the research and design contexts they are being deployed in, echoing the calls among Disability scholars in HCI of \textit{`Nothing About Us Without Us'}~\cite{mankoff2010disability, williams2019nothing, spiel2020nothing}. 

Existing examples of this ``For Us By Us''~\cite{braswell2020FUBU} approach within the Black experience include things like the curation and moderation of spaces like \#BlackTwitter~\cite{klassen2021more, brock2020distributed}, the utilization of the mobile payment platform Cash App as a means to create cultural capitol~\cite{cunningham2022cost}, and even the 2008 launch of the internet browser Blackbird~\cite{brock2011beyond}. In particular, most research focused on the curation of online spaces such as \#BlackTwitter have found that this entity was a place for communities to share content that directly tied into their lived experiences. In contrast, community adoption of spaces and tools that were designed \textit{for} individuals without their input has typically been low. Farnsworth suggests that feelings of belonging, acknowledgment of contributions and healthy environment, and meaningful moderation of these environments can have positive impacts on marginalized groups~\cite{farnsworth2016diversity}, factors which may be better incorporated when communities have input and leadership in the design of these spaces and tools. For example, Dosono et al's work outlines the importance of meaningful moderation in their study of Asian American and Pacific Islander Reddit moderators; indicating that emotional labor is a potential outcome of racially minoritized groups occupying online spaces that are not considered a safe space for communities~\cite{hanckel2019s, dosono2019moderation}.  

Prior research suggests there is value in understanding what it means to design for the cultural and experiential needs of racially minoritized groups. While work has been done around highlighting Black technologists' experiences designing tools for their communities, we still do not have an aggregate definition of designing for lived Black experiences and the role technology plays in these experiences.

%% file: Sections/3_Method.tex
\section{Methodology}

\subsection{Study Overview}
For this study, our goal was to highlight the experiences of Black technologists designing projects that center around the \LBE. To better contextualize what designing for lived Black experiences looks like, we gathered Black technologists' perspectives guided by questions such as: \textit{``What is the `Black experience', how would you define that?"}; \textit{``What is technology’s role in supporting, cultivating, or creating the Black experience?"}; \textit{``What does it mean to you to design for the Black experience?"}; and \textit{``Why do you think you were the best person to implement or take on this project?"} Throughout our study protocol and findings we use the term ``Black experience'' and note that the use of this terminology is meant to capture the individual lived experiences of the Black technologists interviewed in this study. While Black identity is a dimension that has been engaged with in design as a discipline more broadly~\cite{berry2022black}, what the Black experience means differs across domain contexts, supporting the need for our participants' perspectives~\cite{harrington2021designing, bray2021speculative, cunningham2022grounds}.

\subsection{Recruitment \& Data Collection}

Our research study was approved by a university Institutional Review Board prior to our interview studies. To find projects that would qualify for our study, we posted recruitment fliers on social media websites like Twitter and LinkedIn, using the hashtags: \#BlackTech, \#BlackInTech,  and \#Research among others. We included a landing page for our research project detailing prior work in addition to outlining the goals of the study. We scoped our recruitment efforts to the United States, though we were open to relevant projects by technologists in other countries.

We sorted through relevant social media threads to determine if it was a project aimed at supporting the Black community. We determined this by examining whether projects were aimed at supporting the Black community via the target audience of the project or whether or not the product/service/tool sought to uplift or support Black communities explicitly by its description. We then added relevant projects to a spreadsheet that was used to contact participants for interviews. We collected projects for approximately 2 years, starting in August 2020 ~\cite{lisa_egede_2024_11111688}.

Each participant was asked to complete a pre-interview sign-up form where we collected consent forms and background information. Upon receiving completed forms, participants were then scheduled for an interview with at least two members of our research team to ensure no bias based on prior project knowledge. Each interview lasted approximately 1 hour and was conducted by two members of the research team. 
Each interview began with collecting verbal participant consent before asking questions about the relationship between the ``Black experience'' and technology, including the central question \textit{``What is the ‘Black experience’, how would you define that?"} These questions provided a baseline understanding of the wide variety of ways that participants contextualize this concept, as well as any overlapping themes that would help to define it. Then we asked participants to describe their project and detail their motivation behind taking on this project. The entire interview protocol is available online ~\cite{lisa_egede_2024_11111688}.
At the end of each interview, we asked participants to recommend additional people who could be a good fit for the study and augmented our project spreadsheet accordingly. We also asked participants if they wanted to be a part of a publicly available database of people and their projects from the study to increase visibility and promote the amplification of their work. All participants consented to be included in this public database. We compensated each participant with a \$40 Amazon.com gift card.

\subsection{Participant Overview}

We interviewed a total of 17 participants from a database of 48 projects. Of those 17 participants, we recruited 7 Black women, 7 Black men, and 3 Perceptible non-Black (PNB) participants (1 Asian man, 1 White woman, and 1 White man). Given the scope of this study and the goal of each of our research questions, we removed perceptible non-Black participants from the data we analyzed for RQ1 and RQ2. We list all participants, along with a description of their project in Table~\ref{tab:project-description}. 
All but one of our participants were located in the United States at the time of our study. Participant's projects ranged from coloring books to highlight Black women in STEM, to mobile applications that explicitly support the Black community in making connections. Most of our participants had some technical background, though most did not have a degree in computing.

We categorized the types of projects identified as a resource(\faBook), tool(\faWrench), service(\faStore), and community(\faPeopleArrows). We define a resource as a project that provides access to information on a specific topic. We define a tool as a project that allows a user to carry out a specific task. We define a service as a project that provides assistance to a user for a particular task. Lastly, we define community as a project that provides a space, whether physically or digitally, for people with intersecting identities and/or interests to convene.

\input{tables/project-summary-table.tex}

\subsection{Data Analysis}
We analyzed our interviews using a mixed methods coding approach that was both deductive and inductive~\cite{miles2018qualitative}. We used Subply to transcribe the interviews and collaboratively coded the transcripts using ATLAS.ti. We began the analysis with a structural coding process, an approach that consisted of us developing an initial codebook based on our interview protocol~\cite{saldana2009coding}. For example, the rapport-building portion of the interview maps to the \textit{background} category, and the question ``What do you do for a living?'' maps to the \textit{background-jobrole} code.  Once we had our initial codebook, we created coding assignments amongst the research team. Each transcript was deductively coded by two members of the research team. This was intentionally done to ensure diversity of code construction and to make sure researchers were not coding interviews that they conducted. Once each transcript got its first pass by both coders, we collectively discussed the labeling and any additional codes that emerged. Given that additional codes were being introduced throughout data analysis our coding process was inductive~\cite{miles2018qualitative}. Rather than centering our discussions on disagreements regarding what something was labeled, we focused on making sure the descriptions for new and existing codes were clear, concise, and distinct. We chose this methodology to be inclusive of and give due consideration to all interpretations of our participants' insights. The final set of categories and codes from our analysis are publicly available~\cite{lisa_egede_2024_11111688}
\href{https://doi.org/10.5281/zenodo.11122187}{\includegraphics[width=3cm,trim=0 1mm 0 0]{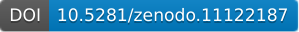}}.

To answer RQ1 and RQ2, we extracted data from the \textit{define} category in our codebook, which maps to data on how participants define the relationship between technology and lived Black experiences. Specifically, for RQ1 we focused on the \textit{define-techblackexp} code, which maps to how participants define technology's role in supporting lived Black experiences, and for RQ2 we focused on the \textit{define-designblackexp} code, which maps to how they define the process of designing technology for lived Black experiences. We also referenced data behind the \textit{define-blackexp} code to help provide context, when necessary, on what ``Black experience'' participants are centering their responses in. Unlike the data needed to answer RQ1, the data for RQ3 did not correspond directly to one or more specific codes or categories. We specifically included perspectives from Perceptible non-Black (PNB) participants in our analysis for RQ3. We conducted a thematic analysis where the first two authors each took a pass to determine the emergent themes in the subset data. We finalized the emergent themes following discussions with all authors.

\subsection{Researcher Statement of Positionality}
Positionality in research refers to where one stands in relation to the topic they're studying~\cite{merriam2001power, bourke2014positionality}.
Based on prior work establishing the value of positionality in research that engages with historically marginalized groups~\cite{cambo2022model}, we find it relevant to acknowledge that our research team was compromised of people who identify as Black or African-American and are all situated within the United States. All research team members are either academic or industrial researchers familiar with HCI research and are college-educated.

\subsection{Limitations} 

Despite recruitment efforts, our current interview sample does not have representation from projects that aim to support Black LGBTQIA+ or Black disabled communities. Participant disclosure of their sexual orientation was voluntary. Upon analysis of our data, we noticed that all of our participants were college-educated, which we observed as a potential limitation. While our study is constrained to the United States/Canada, we note that if we collected data from technologists in other countries there may be additional perspectives on concepts of the lived Black experience and technology's role therein.

%% file: tables/project-summary-table.tex
\begin{table*}[ht]
\centering
\caption{\centering Description of Black (P) \& Non-Black Technologists (PNB) with Project Type.\newline
\faBook = Resource, \faWrench = Tool, \faPeopleArrows = Community, \faStore = Service}
\label{tab:project-description}
\begin{tabular}{l l c}
\toprule
\textbf{PID}             & \textbf{Project Description}       & \textbf{Project Type}                 \\
\midrule
{\participant{P1}}
		& {Data science content from police data on traffic stops}		
		& \faBook  \\
	
{\participant{P2}} & {Black women in STEM coloring book}                                     & {\faBook} \\ 
{\participant{P4}} & {Social network for Black Gen Z professionals}                                                                      & {\faPeopleArrows} \\ 
{\participant{P5}} & {Hub for data \& insights on Black innovators and rising innovations}              & {\faBook, \faStore} \\ 
{\participant{P6}} & {Organization that aids in record expungement of the previously incarcerated}                                                                      & {\faStore} \\ 
{\participant{P8}} & {Mentorship program for college freshmen}                               & {\faPeopleArrows} \\ 
{\participant{P9}} & { 
Professional community-building platform}                            & {\faPeopleArrows} \\ 
{\participant{P10}} & {Mentorship program for students at various HBCUs}                      & {\faPeopleArrows} \\ 
{\participant{P11}} & {Online safe space for Black women and femmes to dream, design, and share experiences }                                                                      & {\faPeopleArrows} \\ 
{\participant{P12}} & {Dating application that matches individuals with professional Black women}                                            & {\faStore, \faWrench} \\ 
{\participant{P13}} & {Platform to support the recruitment and retention of Women of Color in STEM careers}                                                                      & {\faWrench, \faBook} \\ 
{\participant{P15}} & {Website to inform public of missing Black people}               & {\faBook, \faPeopleArrows} \\ 
{\participant{P16}} & {Safety app to track and record interactions from mobile devices} & {\faWrench} \\ 
{\participant{P17}} & {Space for Black folks to dream up culturally relevant tech futures}    & {\faPeopleArrows} \\ 
{\participant{P3}} & {Tool to contact elected representatives about keeping Black lives safe}    & {\faWrench} \\ 
{\participant{P7}} & {Tool to physically share resources to support Black lives}    & {\faWrench} \\ 
{\participant{P14}} & {Tool to make petition-based record clearance equitable}    & {\faBook} \\	  
\bottomrule
\end{tabular}
\end{table*}

%% file: Sections/4_Findings.tex
\section{Findings}
We conducted this study to understand how Black technologists use, create, or curate resources and tools to support the \LBE. We highlight the following themes that emerged from our interviews: \emph{Technology's role in supporting lived Black experiences}; \emph{Motivations and experiences of technologists who design for their \LBE}; and \emph{Influential factors that impact technologists' ability to build out their projects}.

\subsection{Technology's Role in Supporting Lived Black Experiences}
\label{sec:result-rq1}
To address \textbf{RQ1}, we asked participants about technology's role in supporting lived Black experiences.
Participants identified three key perspectives: 1 - The ability to give or deny \emph{access}; 2 - The power to provide \emph{visibility}; and 3 - The affordance to consider innovative futures. Additionally, some participants shared their perspectives on technology's potential for harm.  

\subsubsection{Technology Gives Access} Some participants believed that technology’s role in the \LBE is to give individuals access to create or curate tools, resources, and information for other members of the Black community to leverage. \participant{P8} details the importance that access to data means for their community and the role that technology plays when data is available and someone \inlinequote{creates some sort of tool for us [Black community] to be able to leverage that data so that we can kind of make our lives a little bit easier.} They go on to point out that historically, Black people and certain communities in society have not had access to technology or other resources, thus the ability for technology to \textit{create} access is seen as pivotal. \participant{P11} cites this historical lack of access as a reason for their desire to use technology ``for good'', given that their primary audience is individuals who might not be well-versed with how to use tech. From our participants' perspectives, we see that technology has the power to provide a means of entry, oftentimes correcting the historical lack of access caused by systemic injustices.

\subsubsection{Technology Supports Visibility} Some of our participants shared that technology's role is to provide \emph{visibility} for Black folks and communities through awareness of the \LBE and increasing representation. \participant{P8} reflected on the growth in technology usage stating that as more people become reliant on it, \inlinequote{it's important to also use those outlets to emphasize the Black experience more}, suggesting that technology has the capability to create visibility by way of sharing common experiences. \participant{P4} expands on this idea by stating: \inlinequote{I think the more Black faces that you see in tech, the more you'll see the adoption in underrepresented areas}, emphasizing that the use of technology in Black communities is correlated to the amount of representation in the tech field. For technologists in our study technology is both impacted by and promotes the visibility of marginalized cultures and experiences.

\subsubsection{Technology Provides Opportunity for Futuring and Innovation} \label{tech-future}
Another common thread among our participants was the discussion of how technology can be used as a means of futuring and innovation in the Black experience. \participant{P13} described the \inlinequote{strong points} of the Black experience in relation to tech as \inlinequote{the connection, the ability to imagine and really develop something years ahead of other people.} They go on to describe technology development as a way for Black people to build tools and resources that are ``ahead of their time.'' In regards to diversity in these spaces, \participant{P5} felt that technology had the potential to make a huge impact in increasing representation but \inlinequote{it's not reaching most people as it could be.} Based on these perspectives, technology seemed to operate as a tool for technologists to create innovative experiences for their communities, and to be architects of that future innovation~\cite{harrington2021eliciting, bray2022radical}.

\subsubsection{Technology's Potential for Harm} While many participants focused their responses on the role they believe technology plays in the \LBE, some also discussed the role technology should \textit{not} play, including potentially being harmful or biased. For example, \participant{P15} believed that technology could be a form of deception, stating that \inlinequote{[technology] gives people hope that things are changing.} They go on to suggest that technology may mask or give false perceptions of core problems, further limiting the ability for true change to occur. It was also noted by \participant{P11} that technology can be exclusionary. They go on to state: \inlinequote{I believe in technology, I love technology, but I also think because we often haven't been involved in the creation of those technologies, those technologies haven’t always done right by us.} The historical ``one size fits all'' approach to designing for Black communities~\cite{cunningham2022grounds} has had harmful impacts, as seen by products like the Pulse Oximeter, and the high error rate it had amongst Black users~\cite{rabin2020pulse}. Overall, even though technology has not commonly included or centered Black Americans~\cite{walton_1999, raceinhci}, Black participants in our study found ways to create technology that supported their needs.

\subsection{Designing for Lived Black Experiences}
\input{tables/design-for-summary}

To address \textbf{RQ2} we wanted to understand how technologists design for lived Black experiences. We found that participants described designing for lived Black experiences as multi-faceted. Building on the perspectives highlighted in Section~\ref{sec:result-rq1} we explored technologists' experiences designing for the \LBE. 
Table~\ref{tab:design-for-summary} outlines the core categories of what it means to design for the \LBE that emerged from our discussions with participants: \textsc{Designing for Celebration \& Joy}, \textsc{Designing for Connection}, \textsc{Designing for Culture}, and \textsc{Designing for Continuity \& Survival}. In this table, we highlight the core category each project centered its goals upon with some projects having impact across categories.

\subsubsection{Designing for Celebration \& Joy} \label{sec:celebration-joy}
Participants were motivated by both the desire to develop products that centered community needs and the urge to celebrate Black joy as a result of navigating microaggressions across various aspects of their own lives. When designing their project, \participant{P2} was motivated by their own experiences, noting that when transitioning into computing, \inlinequote{I felt like everyone was coding since the womb, and I was like, I don't even know what's going on.} \participant{P2} shares how isolating it is to be the “only one”, particularly in computer science and other technical majors and career fields. They go on to elaborate on the connection between their motivation and what they see as success for their project, which targets increased representation for young Black girls in STEM:

\blockquo{I feel like it's been successful and not in terms of sales or anything like that, but just the feedback that I've received and just hearing from the students first and foremost. That's what warms my heart, and that's the motivation for me to keep going, numbers aside, money aside, that's the main reason why I did it.}{P2}

For \participant{P2}, their measurement of success is directly tied to the joy of their target audience; the potential to change the narrative or possibly the trajectory of young Black girls' lives meant more than quantitative measures. Similarly, \participant{P5}, whose project centers on amplifying Black innovators, described their efforts as a \inlinequote{a labor of love.} Having had the opportunity to see and experience Black innovation and representation, \participant{P5} became curious as to why these things were not being discussed and celebrated universally:

\blockquo{I came from a place where all of my instructors, the engineers I knew, the software developers, the testers, everyone was Black or Brown in some capacity, so it didn't make sense to me why we were always only asking white guys what they thought, so I wanted that to change.}{P5}

For \participant{P5}'s project, amplification served as a way to celebrate the work of Black innovators as well as to provide greater visibility and economic resources for their work. Centering on the idea of \inlinequote{lifting as you climb,} \participant{P5} shared that they have had the ability to hire Black folks to work for their company or as vendors. In addition to creating a platform that centers around celebrating Black innovation, their hiring of Black employees helps to provide further opportunities and economic support for members of the Black community, tying into prior work outlining the importance of financial support in community-based design~\cite{tang2022community, cunningham2022cost}. Overall, the joy and celebration that resulted from engagement with participants' projects served as both motivation and metrics of success. 

\subsubsection{Designing for Connection} %
Another theme that emerged across participant interviews was that designing for the \LBE also meant designing for connection. As a result of a history of erasure in technological spaces and society more broadly, Black people often find themselves adapting the use of existing tools to meaningfully center their needs~\cite{klassen2021more, brock2020distributed}. For technologists, this meant creating projects to foster community and connect Black folks to tools, resources, or one another.
For example, \participant{P9}, whose project was designed to create a safe space for Black people who might be new to a city or have a business they are trying to grow, stated that they \inlinequote{try to use technology to create a more consistent and high-quality Black experience.} \participant{P9} also discussed the origins and original motivation behind their project.

\blockquo{I created a group chat and set some very simple parameters around hey, let's just talk about events. Let's share information, let's not make it a joking space or a meme space. It's a safe space where we can just share information and you can ask questions.}{P9}

Similar to the concept of \textit{Designing for Celebration and Joy}, \textit{Designing for Connection} speaks to creating spaces for Black people to communicate with one another based on \inlinequote{trustworthy and transparent exchange.} For \participant{P9}, because Black people's lived experiences differ across the board there is a need for data insights on them, as well as insights on how to properly design for Black communities. Similarly, when drawing on their experience designing a dating app for Black women, \participant{P12} cited the lack of representation and inherent bias on existing dating apps as motivation:

\blockquo{If you think about it your algorithm is going to be as biased as the people creating your algorithm right? And so if everything that's feeding into it is, you know, owned by a white male or white woman, which is 99\% of all dating apps on the market, then of course you start to get inherent bias.}{P12}

For \participant{P17}, the importance of designing for the Black experience and ensuring that audiences stay connected meant ensuring Black people doing the work were financially compensated. Or as they (\participant{P17}) put plainly, \inlinequote{pay them [Black people] valuably. Pay them like you would pay your mama, or your guardians.}  
These insights emphasize prior work on how Black people share knowledge via community and how proper compensation for their creativity/ideas has been historically~\cite{klassen2021more} and more recently neglected on popular platforms like TikTok~\cite{johnson2021copyrighting, ile2022black}. For participants in our study, designing for connection involved cultivating a safe space for innovative ideas to grow within the Black community while ensuring the right resources were utilized to sustain their communities.

\subsubsection{Designing for Continuity \& Survival} \label{sec:continuity-survival}

The overall awareness that current systems and resources are not culturally tailored to the experiences of Black folks has pushed technologists to design for continuity and survival based on their lived and shared experiences~\cite{benjamin2019race}. Drawing on their personal experiences, participants shared sentiments of wanting the people in their communities to feel safe and protected in their everyday lives. 

\participant{P15} expressed that their desire to design for the community they grew up in motivated them to create their platform for locating missing Black people. In particular, \participant{P15} spoke about how current media and journalism do not meaningfully engage with the Black community and that \inlinequote{although the resources are good, although the services [are] there, you never reach our communities because it's not culturally appropriate.} According to \participant{P15}'s observations, mainstream news stations' job is to make money, and that involves framing missing people's stories around trauma. \participant{P15} shared that their motivation for building their project was to push people to \inlinequote{take practical steps to ensure the safety of Black people and the safety of Black lives.} Similarly, \participant{P6} discussed how reading about the effects of mass incarceration motivated their project, stating \inlinequote{I can't just be mad that the systems are the systems, I need to figure out how I can change the systems.} This inspired \participant{P6} to pursue their project as a way to address systems of oppression that affect the sustainability of the Black community.

We observed that designing for continuity and survival also meant helping create capacity for Black folks to indulge in various aspects of the \LBE. For \participant{P16}, there was a desire to focus on the joyous aspects of living as a Black person but existing inequalities prohibited them from fully doing so.

\blockquo{If you can't protect and preserve [and] secure your life, then freedom and fulfillment are byproducts of your first living. Tamir Rice was outside a plane, he was gunned down and he never learned how to drive a car. Trayvon Martin, 17. Emmett Till was 14. You look at the ages and our lives are cut short and we don't make it to these milestones of learning how to ride our bikes sometimes or how to drive a car... And all those beautiful, beautiful treasured life moments that just get stolen from us, and the pain that is in the loss of that theft of our lives and our joy and our peace and our fulfillment and our purpose is to be, it's terribly greedy.}{P16}

For \participant{P16}, who designed a security system marketed for Black people, their goal was to protect the lives of young Black children against racially motivated attacks and police brutality; especially with the rise of ``neighborhood watch'' apps like Nextdoor and the use of social media as a monitoring tool more broadly~\cite{bloch2022aversive, bosley2022healing}. For technologists like \participant{P16}, designing for continuity is a demand for the safety of Black people. By creating technology centered around survival, \participant{P16} was also showing their community that \inlinequote{their life is worth something.} Tying into prior literature around the use of Afrofuturism as a means to \inlinequote{make the Black voice central in the design narrative}~\cite{winchester2018afrofuturism}, \participant{P16} expressed that what they have built is meant to save and serve generations to come, further emphasizing this idea that designing for continuity can be thought of as an act that extends to future generations.

\subsubsection{Designing for Culture}

Our participants shared a desire to create technology that pushed beyond themes of ``plight" and instead focused on centering on experiences that collectively tie Black people together. For Black communities, the focus on preserving and amplifying culture through various domains has always been utilized as noted by \participant{P4} who credited the Black community for \inlinequote{integrating [culture] into technology and [making] tech fun and something that's interesting for people to consume.} The use of platforms like TikTok and Twitter were cited as examples, which is a similar observation made by \participant{P9} who drew on the history of Black people using tools to create experiences:

\blockquo{The thing that is consistent with Black people throughout history, is that if you give us tools, we're gonna create culture. Whether it's instruments, whether it's social media. TikTok is a thing because of Black people. Instagram is a thing because of Black people. But they're just tools at the end of the day. It's all about how you leverage them.}{P9}

For \participant{P9}, it was important to acknowledge the innovative contributions Black people have made on social media platforms via popular tech applications~\cite{johnson2021copyrighting, ile2022black}. In response to the lack of representation in the artificial intelligence space, and drawing on their own experiences, \participant{P5} pointed out that \inlinequote{including more Black and Brown people in the structure of the company} could increase representation. In the context of designing for culture in tech spaces, \participant{P5} added that \inlinequote{designing for the Black experience also has to be really swaggy.} He went on to add that \inlinequote{Black folks are the culture}, and this can have profound impacts on tech and design-focused fields. Participants believe that while Black people may have shared experiences, the Black experience is not a \emph{single} experience. \participant{P8} shared: \textit{``I don't really think that there's like a concrete definition because every Black experience is different."} \participant{P9} reinforced this by stating that the Black experience is \textit{``not a monolith."} After \participant{P13} shared the stressful aspects of being of the Black experience, they went on to state: \textit{``at the same time it can be very rewarding because we are a creative people."} This idea of creativity also showed up for \participant{P4} when thinking about how incorporating one's culture into technology can be impactful: 

\blockquo{I think, especially with the Black community we’re more focused on culture, which is good and we know how to integrate that into technology and kind of make tech fun and something that's interesting for people to consume.}{P4}

A similar sentiment was shared by \participant{P11}, who stated that while they love and believe in technology as a tool for sharing culture, it has not always centered Black people in the process. This participant shared that instead, Black communities have had to build their own culture via and around technology, \inlinequote{we are who we are, we've always figured out ways to make [technology] work for us.} This ties into opinions expressed by \participant{P17} who stated that because the Black community has \inlinequote{always longed for African dysphoric relationship} there is a level of creativity that the Black community has \inlinequote{that one generates when you have scarcity.}

While technologists in our study talked about utilizing existing tools for their needs, there exist challenges for shifting the current narrative of what Black culture means in tech. On the idea that culture is centered around trauma and hardships, \participant{P5} talked about the challenges that come with navigating the journalism space as a Black person, such as obtaining funding in their domain space and the expectation to write about \inlinequote{poverty porn narratives.} In this case, designing for the culture meant pushing back on these expectations and thus challenging what the majority interprets or expects Black culture to be.

\subsection{Lived Experiences and Resources: Factors Impacting Development of Technologist Projects}

To answer \textbf{RQ3}, we asked participants about their experience designing their project and to describe what impacted their ability to build them out. Among these questions included: why are you the best person to take on this project? Responses ranged from noticing a lack of representation to the realization that they possessed the resources to take on the work. Black technologists cited reasons such as filling a gap, which differed from that of non-Black technologists in our study who listed having the time and resources as reasons to take on their projects. Furthermore, we found that the rationale provided by one group was often perceived as a challenge for the other (e.g., Black technologists not having the time and resources of their non-Black counterparts). In this section, we outline the themes that emerged from discussions of what participants felt made them most qualified to create their projects and what meaningfully building these projects looked like for technologists more broadly.

\hfill

\subsubsection{Lived Experiences as Motivation}~\label{sec:whybest-livedexp} For many of the Black technologists we interviewed, their lived experiences contributed to them being the ``best person'' to lead their project. One aspect of the \LBE that emerged as a prominent theme in these responses was the lack of representation in existing technologies and resources. \participant{P4} outlined that their motivation to take on their project stemmed from their desire to be a figure that young Black kids can look up to, especially in a space like the tech startup world:

\blockquo{Like I know a decent amount of Black founders only because I'm just in that bubble, but a lot of people aren't and they don't ever see the process. They don't ever know, you know, like what people are building, why they're doing what they're doing.}{P4}

\participant{P4} felt they were the best person because of their resilience and that it takes a lot of work to build projects like theirs from the ground up. 
When thinking about what it meant to be the best person for their project, \participant{P5} acknowledged that the project needs evolved. Because of this, they stated that while they think they are the best person to lead this now, they may not be forever:

\blockquo{I think there are so many other entities that will be able to do this work even better as it becomes much more professionalized, and I'm looking forward to that. My time horizon for [redacted] is like three years before we get purchased and get institutionalized at a larger entity that can do this work better, faster, and cheaper with some of the most talented and brilliant minds.}{P5}

For \participant{P5}, thinking about how their role impacts the communities they are designing for informs how they envision the future of their technology. 
Similarly, \participant{P1} began creating resources for Black youth as a result of observed gaps in the tech space, stating that \inlinequote{if I didn't see it, I was going to do it myself.} If they saw online resources that would benefit their peers or community they would share the resource, which also served as a means of amplification for content creators.

For technologists in our sample, curating and sharing resources is one way to increase representation and create safe spaces with and for their communities. Designing these spaces for ``everybody'' versus designing spaces for the Black community was a point of contention for \participant{P11}. After spending years working in non-minoritized spaces, \participant{P11} realized their creativity, along with the creativity of their Black peers, was being stifled. They elaborated on this idea of being in a \inlinequote{constant state of survival mode,} stating:

\blockquo{[It] just got me to the point where I felt like I was tired of designing spaces for everybody but us and I wanted to design the space that I would wanna work and create things. Where would I feel safe or where would I have a cheering squad that when I come with the idea for how we could be freer and I say, `Hey, I wanna try this out,' they're not gonna look at me like I'm crazy or tell me it's impossible or give me a billion excuses as to why it can't happen.}{P11}

Overall, technologist’s struggles navigating their respective spaces translated into how they approached the technology they were creating for their communities. They were often trying to fill a gap, whether that be increasing representation in their respective fields or sharing resources to benefit their communities. Additionally, the support, expertise level, and lived experiences of our participants were an integral part of their ability to push their projects forward.

\subsubsection{Resource Availability \& Access}

In regards to resource availability and access, we found that Black and non-Black technologists had different experiences, which had an impact on their ability to contribute to their projects. Following the Black Lives Matter (BLM) movement in 2020, there was a rise in ``tech for social good" tools centered around uplifting the Black community and other marginalized groups impacted by racism~\cite{hankerson2016does, green2019good}. This movement paired with the COVID-19 pandemic gave some technologists free time that they might not have had otherwise. Non-Black technologist \participant{P3} stated they created a tool that allows people to see which politicians represent them in their district. When we asked why they were the best person they stated: \inlinequote{I'm not the best person but I am the person.} \participant{P3} went on to state that being from a smaller town it is hard to find tech that is tailored to their needs, which led them to make a tool to simplify finding local representatives. Although they did not describe themselves as the best person to take on the project, they stated that they \inlinequote{had a lot of free time 'cause of the pandemic} and figured that this was something they wanted to work on while trapped at home.

For some participants, having support from family doubled as a resource as it helped free up time for their projects. 
Black technologist \participant{P13} stated that they did not know if they were the best person but having family support made it easier to fully focus on their project. For \participant{P13}, it wasn't about being the most qualified person, but \inlinequote{a person who has a support system and resources that has allowed [them] to make it to year five in a startup.} This is similar to the sentiments felt by Black technologist \participant{P16} who, when citing their dad as motivation for getting things done, stated that \inlinequote{the person who may be the least qualified but actually gets it done, was ultimately the best person to do it.}

Our non-Black participants cited having access to resources in the form of voluntary support and domain knowledge from local activists, such as \participant{P3} who spoke about the feedback they got from Black local activists:

\blockquo{A lot of my mutuals and followers were also [redacted] students. So there was a bunch of Black groups at [redacted] who reached out to me about it. And had input and feedback on how to draft or edit some stuff. Some changes like the advocating for abolition were things that I didn't end up implementing, but it was still really cool because I don't think I was ever that involved in that community before.}{P3}

For \participant{P3}, local activists and peers were proactive about reaching out to see where they could lend their resources. However, lack of domain knowledge meant unintentionally leaving Black activists out of the conversation when it came to their tech design, as observed by the experience of non-Black technologist \participant{P7}:

\blockquo{The local Black Lives Matter (BLM) chapter reached out to us. The head of that chapter was like upset. She was like ``How did you not include us in this conversation?'' Then we didn't know about this and we were totally taken aback. I didn't even know until then that BLM had local chapters and it had municipal-level organizations and we were apologetic. We were like we're so sorry we didn't mean to not include you. We didn't know you were here.}{P7}

For \participant{P7} and their team, unintentionally leaving out expertise from activists and community organizers can be attributed to the level of existing knowledge they had about the communities they were designing for. Being connected with local activists had a positive impact on the type of feedback that non-Black technologists received on their projects, but for \participant{P7} there were still reservations when it came to how their technology would be perceived by their audiences. When asked why they were the best person to implement their project, P7 stated they were not the best and even discussed their fear of receiving backlash:

\blockquo{We were actually pretty nervous [...]. We were three white founders. White dudes. And we were like kind of nervous about- Should we even try this? Because is it gonna be perceived wrong, are people gonna think we're just capitalizing on something and trying to get attention> Anyway, there were a lot of nerves we didn't think we were the best people, but at the end of the day we thought it's worth people having this and talking about it. More than it is us worrying about backlash. I'd rather have a conversation and learn and mess up, then be scared and not.}{P7}

While available time and resources contributed to the ability of non-Black technologists in our study to pursue their ideas, some cited other reasons they felt they were the best people for their project. \participant{P3} cited the desire to make a positive impact on the Black community as motivation. A similar sentiment was felt by non-Black technologist \participant{P14}, who expressed that their goal was to take the burden off of communities most impacted by racial disparities.

Overall, non-Black technologists' motivations centered around offering support to the Black community, but lack of community-specific knowledge resulted in them unintentionally leaving Black activists and organizers out of their design process. When it comes to designing tools that are being created as a means of survival, technologists with lived experience had more domain expertise, which played a critical role in how they built their projects. Additionally, our findings support this idea that designing for the \LBE requires financial resources and labor; some of which appeared to be more readily accessible to non-Black technologists. For Black technologists conducting this work, there was the acknowledgment that the playing field is not leveled~\cite{equale} and that there will be extra hurdles to overcome to ensure that tech-centering Black people reach their fullest potential.

%% file: tables/design-for-summary.tex
\begin{table*}[ht]
\centering
\caption{Core Categories of What it Means to Design for the Black Experience.}
\label{tab:design-for-summary}
\begin{tabular}{c l l}
\toprule
\textbf{Core Category} & \textbf{Category Description˛} & \textbf{Aligned Projects}                 \\
\midrule
\multirow{4}{*}{\parbox{2.5cm}{\centering \textsc{Designing for Celebration \&Joy}}}	
		& 		
		&    \\
		
		& \begin{tabular}{@{}l@{}}Projects centered around happiness and\\ uplifting of the Black community \end{tabular}&  \participant{P2}, \participant{P5}  \\ 

		& &    \\

	    \midrule
		\multirow{4}{*}{\parbox{2.5cm}{\centering \textsc{Designing for Connection}}}
		&  
		&  \\ 
		
		& \begin{tabular}{@{}l@{}}Projects focused on creating and fostering\\ communities of Black people by connecting \\individuals to others or specific resources \end{tabular} & \begin{tabular}{@{}l@{}} \participant{P4}, \participant{P8}, \participant{P9}, \participant{P10}, \\ \participant{P12}, \participant{P13}, \participant{P17}\end{tabular}\\ 
		
		& &  \\

  	    \midrule
		\multirow{4}{*}{\parbox{2.5cm}{\centering \textsc{Designing for Culture}}}
		&  
		&  \\ 
		
		& \begin{tabular}{@{}l@{}}Projects centered around the appreciation \\and amplification of the various aspects of\\  Black culture\end{tabular} & \begin{tabular}{@{}l@{}} \participant{P4}, \participant{P5}, \participant{P9}, \participant{P11} \end{tabular} \\ 
		
		& &  \\

			\midrule
		\multirow{4}{*}{\parbox{2.5cm}{\centering \textsc{Designing for Continuity  \& Survival}}} 
            & 
		&  \\ 
  
		&\begin{tabular}{@{}l@{}} Projects around keeping Black people and\\ their livelihood safe and protected \end{tabular} & \begin{tabular}{@{}l@{}} \participant{P1}, \participant{P6}, \participant{P15}, \participant{P16},\\ \participant{P3}, \participant{P7}, \participant{P14} \end{tabular} \\
		
		& &  \\ 
\bottomrule
\end{tabular}
\end{table*}

%% file: Sections/5_Discussion.tex
\section{Discussion}

Conceptualizing the Black experience as a part of the design process is a topic area that has yet to be explored in HCI, largely due to the lack of representation in tech and the design space more broadly~\cite{ogbonnaya2020critical, himmelsbach2019we, linxen2021weird}. As a result of this historical under-representation, Black technologists have been pushed to this form of design as a means to uplift and re-center their communities~\cite{equale, tran2019gets, ogbonnaya2020critical}. In this study, our participants shared their experiences cultivating technological tools as a resource for their community, what this approach to design meant for them, and their experience as technologists taking on this work. We discuss and interpret our findings in the following sections, and conclude with considerations for engagement for the broader design community.

\subsection{Representation, Belonging, \& Lived Black Experiences} 

Across HCI and design, historically marginalized groups are advocating for direct participation and a voice in research and development that directly impacts their communities~\cite{spiel2020nothing, ogbonnaya2020critical}. Scholars in our field acknowledge that this inclusion and representation does much to attune to both power and bias in technological systems, and thus participation of marginalized groups must be sought across all design activities~\cite{ogbonnaya2020critical}. For example, prior work conducted on how Black people create community through Twitter supports the idea that designing technology that centers culture is a means for liberation~\cite{klassen2021more, brock2020distributed}. The critical role technology plays in the \LBE was expressed by our participants who described the use of technology as an amplification tool that could make their work more visible in addition to centering Black voices in their communities. The shortcomings that can arise from utilizing apps that were not built for \textit{their} Black experience were cited as motivation to design to center themes around culture and joy. Participants' descriptions of the creativity that encompasses Blackness and what it means to design for their community were prevalent, with the influence of Black culture across various mediums cited as support. While the positive aspects of creating tools to center the \LBE were celebrated, the difficulty of navigating technological spaces continued to be a problem area for technologists. In particular, Black culture is shared through social media apps like Twitter and TikTok, having a widespread influence on trends on these and other platforms~\cite{ile2022black}. Black influence on such platforms is widely profound, and seemingly justified, as many non-Black creators have utilized Black influence for monetary gain, sparking protests and backlash from Black creators~\cite{ile2022black}. Monetization of Blackness paired with the lack of support and centering of Blackness through design was noted by participants, especially those who were designing as a means to cultivate and preserve their culture. Based on this, it is understandable why Black technologists might be more intentional about ensuring their community is centered and prioritized in respect to their projects, especially having observed the drawbacks of celebrating Blackness on platforms that are not designed for Black communities.

The observed domino effects from under-representation have had a direct impact on how many Black people can utilize tools to design for their communities. Our participants expressed the lack of representation was felt throughout their lives including in educational spaces and now present once again in technological spaces. Self-preservation and a need for inclusivity within their lived experiences ultimately informed how they approached designing the projects included in our analysis. This was in contrast to the experiences of non-Black participants, whose major barriers included a limited knowledge of the communities they were designing for. While Black participants utilized their friends, family, and communities to build out their projects, they found that building projects in non-minoritized spaces came with barriers, some of which included funding and lack of access to social capital. Tying into the concerns participants expressed around proper compensation for Black creators, funding acted as a roadblock for participants who otherwise had the educational qualifications to navigate tech spaces. While prior work has outlined the utilization of technology to gain social capital in economically distressed communities~\cite{dillahunt2014fostering}, our participants' experiences emphasize the critical role that social capital plays when navigating tech spaces where participants' technical backgrounds are qualified, if not more. Additionally, the need for data that is representative of Black communities was cited as a desired resource, but due to the history of white hegemony in design such resources are scarce~\cite{linxen2021weird}. 

For Black technologists, and particularly those who were designing projects to combat racism and structural inequalities, there was a major focus on survival which limited technologists' existing space to bask in the joy and explore the innovative aspects of designing for their communities. It should also be noted that of the 3 non-Black participants in this study, their projects fell under the ``continuity and survival'' category (Section~\ref{sec:continuity-survival}), raising questions about the scope and capacity that non-Black technologists have when understanding the nuanced needs and aspects of Black communities that might relate to joy and pleasure. Can non-Black technologists truly design and build for the joy, culture, and connection experienced within the Black community without having been informed through lived experiences? 

Given the uniquely multi-faceted nature of Blackness, it can be further emphasized that designing for the lived Black experience is \textit{not} a monolith and that there is value in understanding how different Black communities are designing for the everyday. This may require that consider all aspects of identity at both the individual and community level, and consider Blackness outside of the scope of what is lacking~\cite{cunningham2022grounds}. Our findings suggest that designing for lived Black experiences goes beyond tackling inequities; it centers around themes of happiness and a desire to uplift the Black community through educational tools and resources. Overall, technologists' desire to preserve existing culture within communities motivated them to create tools to connect Black folks to resources and each other. While we contextualize this experience from the perspective of individuals in the United States, we recognize that this opens up possibilities to consider lived Black experiences outside of this context.

\subsection{Considerations for Integrating the Lived Black Experience in the Design Process}
Based on our findings we provide considerations to amplify the needs of technologists who are designing for the \LBE. Additionally, we outline dimensions for the DIS community to consider when supporting Black technologists, some of which include intentional long-term support for technologists and meaningful engagement practices. We also outline how insights from this study can better inform how designers and researchers can approach creating technology for Black communities.

\subsubsection{Directly Support Black Technologists via Funding and Amplification} 
As observed in our findings, what the \LBE meant to participants centered around themes of celebration, connection, culture, and continuity, with each of these elements serving as an informative dimension in our understanding of their design process. We found that participants had a considerable amount of tech expertise and while it might have been assumed that domain-specific knowledge would open doors for participants, Black technologists in our study still faced challenges obtaining resources needed to advance their projects~\cite{klassen2023black}. Not only does this observation highlight the need for acknowledging inequities in design and tech spaces more broadly, but also the importance of finding meaningful ways to support these projects through funding, patronage, or exposure. For the HCI community, this could mean directly collaborating with Black technologists on research, and utilizing or amplifying tools that promote educating or teaching Black communities about tech and related tools. An additional dimension for the design community to consider is the degree to which technologists want their work highlighted. In particular, for technologists who might be working on projects that involve sensitive topics or those that would require they remain anonymous for safety or personal reasons, it may be difficult to know where to draw the line between advocating for the privacy of participants and uplifting their work. In the case where amplification may not be appropriate, funding or resource allocation may be viable ways to support Black technologists. Thus, being intentional about the ways in which we support Black technologists inside and beyond the HCI and design community may make amplification efforts more sustainable for technologists long term.

\subsubsection{Be Intentional About Recruitment and Long-term Engagement of Minoritized Groups} 
Our participants emphasized how strong community-based networks are. Prior literature reveals that the stakes are often higher for minoritized groups, suggesting the importance of sustaining community-based connections and  \emph{meaningful} community engagement~\cite{hardy2019participatory, cunningham2022cost, bray2021speculative, spiel2020nothing, harrington2019deconstructing}. In our study, we created a detailed database that helped track descriptions of technologists' projects and other important data \cite{lisa_egede_2024_11111688}. Using such tools to log detailed information about projects and their creators can help ensure that designers are attuned to the goals of their participants' work and can also serve as a tool to stay up to date on their projects following the conclusion of the study. In particular, keeping participants up to date on the study or design progress and giving them an opportunity to sustain engagement with other projects or opportunities that align with their needs. Additionally, it can push researchers to be more critical about the types of requests they bring to participants and how they are engaging with participants' time. Approaching minoritized groups with intentions centered around understanding their work can help ensure that trust is established between them and researchers. Additionally, providing proper compensation can be considered a form of meaningful long-term engagement, as it can target financial, project, or community-specific needs.  

\subsubsection{Increase Efforts to Share Design-Focused Research Beyond the DIS Community}
Access to education and resources was a prominent theme for technologists, especially as it related to advancing their projects and positively impacting their communities. As HCI researchers and designers, we are often extracting knowledge from communities and synthesizing our findings in a way that is digestible for our own research communities. There is a need for critical thought around how we can increase access to research spaces where this knowledge is being shared and facilitate conversation between designers and technologists, builders, community leaders, and non-academics more broadly. For example, considering how we advocate for more funding opportunities for individuals like the technologists in our study to access spaces such as academic conference venues and university-led talks might be a promising approach. Ensuring that these events are being publicized beyond an internal network can help ensure that non-academics are being included in conversations that directly tie to their lived experiences. Additionally, we can take more steps to ensure that we meaningfully facilitate conversations between researchers and technologists once they are in these spaces by promoting collaborative work and community-centered events. Providing a `seat at the table' is a powerful way for HCI researchers to empower technologists doing this work, furthering our understanding of what it means to design and build for lived Black experiences.

\subsubsection{Design to Create Capacity for Joy in Black Communities}

While extensive research has been done to better understand how to design for minoritized populations, it is often one-dimensional or prefaced by how these communities experience deficits and inequities~\cite{cunningham2022grounds, to2023flourishing}. Many technologists in our study were either designing \textit{in response} to products or experiences that did not meaningfully engage with their communities or designing to offset the harm caused by poorly designed tools. Having seen the negative effects that products had on their communities post-deployment, technologists in our study created projects to not only have long-term impacts but to offset the perceptions of their communities. Given this, we urge designers and researchers to prioritize `intent' when helping build tools or products for minoritized communities. Throughout the design process, we can ask: ``What does it mean to design products that meaningfully engage with and center lived Black experiences?'' Oftentimes harmful products are rooted in good intentions, which is why it is important to focus on the multifaceted needs of communities. Giving communities products that create time and space for them to center the uplifting aspects of their experiences can be thought of as an impactful approach, supporting the idea of ``designing for the capacity to create joy''. \textit{Continuity \& Survival} based projects focused on protecting Black lives could instead be thought of as a way to give Black communities the opportunity to center joy, culture, and connection.

%% file: Sections/6_ConclusionandFutureWork.tex
\section{Conclusion \& Future Work}

 This paper aims to understand how technologists have used, created, or curated resources to support the lived Black experience. Based on our study findings, we found that challenges for Black technologists ranged from inadequate resources to limited social and financial capital. By connecting the Black community to tools needed to thrive as well as to each other, our technologists found designing for lived Black experiences to be multifaceted; with a focus on themes related to community support, the cultivation of culture, survival, and the centering of joy. These findings help to situate that our participants do not merely see Blackness as being a problem to solve, but rather as a unique way of being in a society that has structures in place to oppress and minimize cultural identity. Across various domains, it is important to understand how Black technologists design and develop tools and experiences for their community and the everyday. Given that the goal of this work is to analyze the experiences Black technologists have when designing for lived Black experiences, non-Black participants were not included in the data analyzed for RQ1 and RQ2. We anticipate that future work will include their experiences in data analysis and focus on their challenges around designing for Black experiences.

 Overall, insights from this work emphasize the need for meaningful engagement with Black people and technologists in the design process. We hope that findings from this work can serve as a resource for researchers and designers to better understand what it means to design for the \LBE and emphasize the idea that designing in the margins is enough.